\documentclass[conference]{IEEEtran}

\usepackage{cite}
\usepackage{easybmat}
\usepackage{stfloats}
\usepackage{amsmath,amssymb,amsfonts}
\usepackage{graphicx}
\usepackage{textcomp}
\usepackage[linesnumbered,lined,boxed,commentsnumbered, ruled]{algorithm2e}
\usepackage{epstopdf}
\usepackage{balance}
\usepackage{caption}
\usepackage{subcaption}
\usepackage{algorithmic}
\usepackage{xcolor}
\usepackage[letterpaper, left=0.7in, right=0.7in, bottom=1.1in, top=0.71in]{geometry}

\newtheorem{lem}{Lemma}

\DeclareFontFamily{U}{mathx}{}
\DeclareFontShape{U}{mathx}{m}{n}{ <-> mathx5 }{}
\DeclareSymbolFont{mathx}{U}{mathx}{m}{n}
\DeclareFontSubstitution{U}{mathx}{m}{n}

\DeclareMathAccent{\widecheck}{0}{mathx}{"71}

\newenvironment{skproof}{\paragraph*{Sketch of Proof}}{\hfill$\square$}
\renewcommand{\baselinestretch}{.9} 

\newcommand{\mrm}[1]{\ensuremath{\text{#1}}}
\renewcommand{\vec}{\ensuremath{\mrm{vec}}}

\newcommand{\bs}[1]{\ensuremath{\boldsymbol{#1}}}
\newcommand{\comment}[1]{}

\newtheorem{as}{Assumption}
\renewcommand{\H}{\boldsymbol{H}}

\begin{document}

\title{Channel Orthogonalization in Panel-Based LIS}
\author{Juan Vidal Alegr\'{i}a, Ove Edfors\\
\IEEEauthorblockA{Department of Electrical and Information Technology, Lund University, Lund, Sweden\\} 
juan.vidal\_alegria@eit.lth.se, ove.edfors@eit.lth.se}

\maketitle
\begin{abstract} 
Large intelligent surface (LIS) has gained momentum as a potential 6G-enabling technology that expands the benefits of massive multiple-input multiple-output (MIMO). On the other hand, orthogonal space-division multiplexing (OSDM) may give a promising direction for efficient exploitation of the spatial resources, analogous as what is achieved with orthogonal frequency-division multiplexing (OFDM) in the frequency domain. To this end, we study how to enforce channel orthogonality in a panel-based LIS (P-LIS) scenario. Our proposed method consists of having a subset of active LIS-panels coherently serving a set of users, and another subset of LIS-panels operating in a novel low-power mode by implementing a receive and re-transmit (RRTx) process. This results in an inter-symbol interference (ISI) channel, where we characterize the RRTx processing required to achieve simultaneous orthogonality in time and space. We then employ the remaining degrees of freedom (DoFs) from the orthogonality constraint to minimize the RRTx processing power, where we derive a closed-form global minimizer, allowing for efficient implementation of the proposed scheme.
\end{abstract}

\section{Introduction}\label{section:intro}
Large intellignet surface (LIS) \cite{husha_data} is a technology which extends the concept of massive multiple-input multiple-output (MIMO) \cite{marzetta} by considering whole walls covered with electromagnetic active material serving a set of user equipments (UEs) within the same time-frequency resource. The potential of LIS mainly derives from its increased spatial resolution, allowing to multiplex users not only in the angular domain, but also in the depth domain\cite{husha_data,emil_next}, thus leading to improved spectral efficiency beyond that of massive MIMO.

The main theoretical results on LIS---also found under the term holographic MIMO---have relied upon modeling it as a continuous surface of electromagnetically active material \cite{husha_data}. However, practical implementations of LIS may potentially consist of a discrete set of antennas deployed throughout large surfaces, which has firm equivalence with the continuous counterpart as proved in \cite{husha_data,pizzo}. Moreover, practical deployments may also favor panel-based LIS (P-LIS) approaches, where a LIS is divided into several panels of antennas which may be distributed throughout large areas \cite{andreia,jung}.

As remarked in the literature \cite{cell-free_survey,emil_next,wax_journal}, having distributed panels of antennas coherently serving a set of UEs leads to critical issues in terms of interconnection bandwidth and processing complexity. These issues may be tackled by employing decentralized approaches where the panels may preprocess the received signals before combining them and/or sending them to a central processing unit (CPU) \cite{scalable_cell-free,jesus-lis,wax_journal}. In this work, we propose an alternative approach to relieve these issues which consists of using only a subset of active LIS-panels for jointly decoding the data from the UEs, while the remaining panels would operate in a \textit{low-power mode} with the aim of easing the decoding task for the active panels. Specifically, we consider employing the low-power panels to enforce channel orthogonality in space---as well as time---allowing for orthogonal space-division multiplexing (OSDM).

The idea of OSDM was considered in \cite{rs_orth_journ} as a counterpart to orthogonal frequency-division multiplexing (OFDM) for the spatial domain, which offers a convenient framework for efficient decoding and resource allocation in MIMO-based systems. Initial results from \cite{rs_orth_journ} show how to achieve orthogonal multi-user MIMO channels through the use of passive reconfigurable surfaces, where the freedom in the orthogonality constraint is employed to assure the passive nature of reconfigurable surfaces. In the current work, we however consider the use of active LIS-panels for achieving channel orthogonality by implementing a receive and retransmit (RRTx) process employing a subset of the LIS-panels in low-power mode. The main difference is that, in order to maintain coherence and synchronization using baseband (BB) processing at the LIS-panels, the signals retransmitted through the RRTx process are delayed one time-slot before arriving the active LIS-panels, thus generating an inter-symbol interference (ISI) channel. One important benefit of the current approach over those using passive surfaces \cite{rs_orth_journ,nulling_ris} is that, in OFDM-based systems, the BB processing units (BBU) of the LIS-panels may perform independent processing per narrowband subcarrier, unlike with passive surfaces where optimal performance is degraded throughout subcarriers \cite{RIS_WB}.

The considered orthogonality restriction allows for some degrees of freedom (DoFs), which may be exploited to minimize the power of the RRTx processing, similar to \cite{rs_orth_journ,globecom22} when trying to fulfill the passive constraints. In the current framework, we have even characterized the closed-form global minimizers of the RRTx processing power. This also poses an advancement over \cite{rs_orth_journ,globecom22}, where the solution to the analogous optimization problems employed costly minimization algorithms without assured convergence to a global minimizer.

The rest of the paper is organized as follows. Section~\ref{section:model} presents the system model. In Section~\ref{section:orth}, we show how to perform channel orthogonalization in this framework. In Section~\ref{section:Pmin}, we derive the closed-form solution to the power minimization problem. Section~\ref{section:num_res} includes a numerical evaluation of the proposed methods. Section~\ref{section:conc} concludes the paper with some final remarks.

\section{System Model}\label{section:model}
\begin{figure}
    \centering
    \includegraphics[scale=0.52]{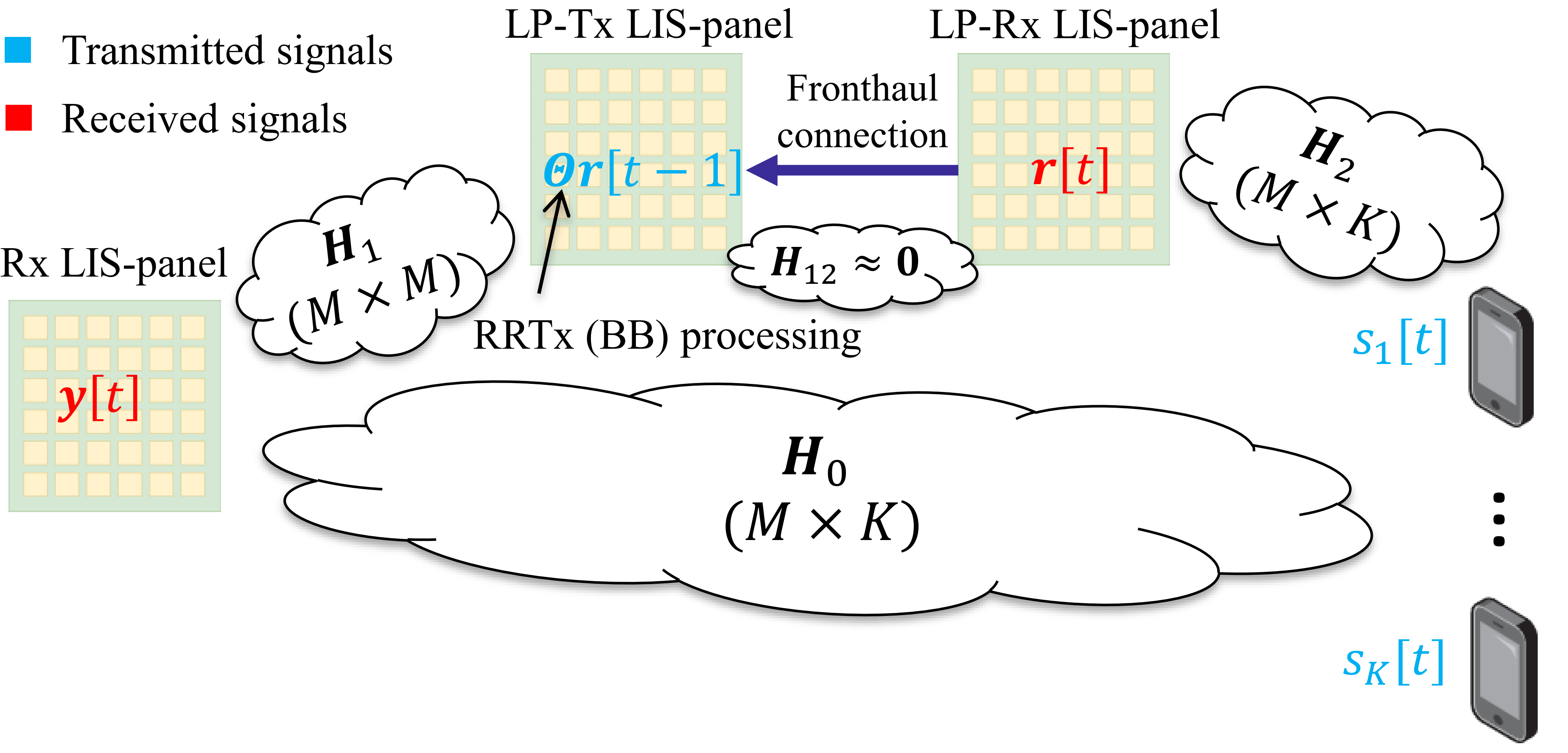}
    \vspace{-0.3em}
    \caption{Illustration of the scenario at time-slot $t$.}
    \label{fig:scenario}
    \vspace{-1.5em}
\end{figure}

Let us consider a P-LIS scenario\footnote{Equivalence with cell-free massive MIMO scenarios is evident by noting the correspondence between a LIS-panel and a distributed multi-antenna AP.} as illustrated in Fig.~\ref{fig:scenario}, where a total of $3$ LIS-panels are used to decode the uplink transmission of $K$ UEs through a narrow-band channel (e.g., one OFDM subcarrier). For simplicity, all LIS-panels are assumed to have the same number of antennas $M\geq K$, but extension to different number of antennas is straightforward. We believe that the $3$-panel scenario captures the essence of the proposed solution, but in the extended version we may formalize the results for an arbitrary number of LIS-panels.

Out of the $3$ LIS-panels, only $1$ panel is actively receiving and decoding the data from the UEs. The $2$ remaining LIS-panels operate in a hereby defined \textit{low-power mode}, where they jointly implement a receive and retransmit (RRTx) process. In this process one of the low-power LIS-panels, the \textit{low-power receive} (LP-Rx) panel, receives the signals being exchanged between the UEs and the active panel. The other panel, termed \textit{low-power transmit} (LP-Tx) panel, simultaneously retransmits, after BB processing, the signals previously received at the LP-Rx panel, which are shared via fronthaul links. Note the need of two panels to allow simultaneous reception and transmission of BB data, but our approach could also employ a single low-power panel with full-duplex radios. In the RRTx process, the low-power panels may also perform linear processing, which may be physically done at the BBU of either (or both) of the panels. The received vector at the active panel at time-slot $t$, is then given by
\begin{equation}\label{eq:ISI_yt}
    \bs{y}[t] = \bs{H}_0\bs{s}[t]+\bs{H}_1\bs{\Theta}\bs{r}[t-1]+\bs{n}[t],
\end{equation}
where $\bs{H}_0$ is the $M\times K$ channel matrix between the UEs and the active panel, $\bs{H}_1$ is the $M\times M$ channel matrix between the LP-Tx panel and the active panel, $\bs{r}[t]$ is the $M\times 1$ received vector at the LP-Rx panel at time-slot $t$, $\bs{\Theta}$ is an $M\times M$ matrix associated to the BB RRTx processing, and $\bs{n}[t]\sim \mathcal{CN}(\mathbf{0},N_0 \mathbf{I}_M)$ models the additive white Gaussian noise (AWGN) added at the active panel at time-slot $t$. We may express $\bs{r}[t]$ recurrently by
\begin{equation}\label{eq:ISI_rt}
    \bs{r}[t] = \bs{H}_2 \bs{s}[t]+\bs{H}_{12}\bs{r}[t-1]+\tilde{\bs{n}}[t],
\end{equation}
where $\bs{H}_2$ is the $M\times K$ channel matrix between the UEs and the LP-Rx panel, $\bs{H}_{12}$ is the $M\times M$ channel matrix between the LP-Tx and the LP-Rx panels, and $\tilde{\bs{n}}[t]\sim \mathcal{CN}(\mathbf{0},\tilde{N}_0 \mathbf{I}_M)$ models the AWGN added at the LP-Rx panel. 

\section{ISI Channel Orthogonalization}\label{section:orth}
We begin this section by considering two simplifying assumptions that will make the problem more tractable, while still covering a reasonable range of scenarios. The validity of the first of these assumptions will be further reinforced by the results from Sections \ref{section:Pmin} and \ref{section:num_res}.
\begin{as}\label{as:noise}
The noise term added in the RRTx process has negligible impact in \eqref{eq:ISI_yt}. Note that this extra noise term appears multiplied by $\bs{\Theta}$ in \eqref{eq:ISI_yt} so it may have non-diagonal covariance---i.e., leading to colored noise. However, we will minimize norm of $\bs{\Theta}$, which will reduce the impact of this noise term compared to the white noise term $\bs{n}[t]$. Thus,the $\tilde{\bs{n}}[t]\approx\bs{n}[t]$ may be assumed approximately white.
\end{as}
\begin{as}\label{as:H12}
Since the panels may be conveniently placed within the scenario, we assume that the wireless channel between the low-power panels is negligible. For example, the SP-Rx panel could be pointed at some indoor UEs, and the the SP-Tx panel could be pointed at an outdoor active panel. We thus assume $\bs{H}_{12}\approx \boldsymbol{0}$ such that we can get rid of the recurrence associated to \eqref{eq:ISI_rt}.\footnote{Note that, even if $\bs{H}_{12}$ is not exactly $\bs{0}$, every step in the recursion would attenuate its contribution due to the repeated path-loss being experienced.} In the case of a single low-power panel with full-duplex operation, this assumption corresponds to having good isolation between transmitter and receiver.
\end{as}

Considering Assumptions \ref{as:noise} and \ref{as:H12}, we may rewrite \eqref{eq:ISI_yt} as
\begin{equation}\label{eq:ISI_yt2}
    \bs{y}[t] = \bs{H}_0\bs{s}[t]+\widetilde{\bs{H}}(\bs{\Theta})\bs{s}[t-1]+\bs{n}[t],
\end{equation}
where the $M\times K$ matrix $\widetilde{\bs{H}}(\bs{\Theta})=\bs{H}_1\bs{\Theta}\bs{H}_2$ corresponds to the aggregate channel associated to the RRTx process. Note that, assuming full-rank $\bs{H}_1$ and $\bs{H}_2$, as well as unrestricted RRTx processing $\bs{\Theta}$, we have full freedom in designing $\widetilde{\bs{H}}(\bs{\Theta})$. For example, this could be achieved by canceling $\bs{H}_2$ in the LP-Rx panel (multiplying the left pseudo-inverse) and then cancelling $\bs{H}_1$ in the LP-Tx panel (multiplying the inverse) after applying the desired processing on the signal coming from the LP-Rx panel. Note that, if Assumption~\ref{as:noise} does not apply, the colored noise term in \eqref{eq:ISI_yt2} would limit the optimality of symbol-by-symbol detection after performing maximum ratio combining (MRC) with the resulting orthogonal channels. However, the results that we show next would still lead to channel orthogonality in the desired signal part of \eqref{eq:ISI_yt2}, i.e., leading to perfect interference cancellation.

As can be seen from \eqref{eq:ISI_yt2}, due to the RRTx process, the received vector suffers from ISI, thus compromising the orthogonality property in the time domain. Considering $T$ consecutive symbol transmissions within the coherence time, and assuming $\bs{s}[t]=\bs{0}$ for $t=0$ and $t=T+1$, we may write the complete received vector as
\begin{equation}\label{eq:ISI_y_vec}
\bs{y}_\mrm{C}=\bs{\mathcal{H}}\bs{s}_\mrm{C}+\bs{n}_\mrm{C},
\end{equation}
where $\bs{y}_\mrm{C}$ is a $(T+1)M\times 1$ vector stacking up the received vectors $\bs{y}[t]$ at $t=1,\dots,T+1$,\footnote{We assume that the receiver may be active for one more time-slot than the transmission duration so that it can listen to the last retransmission. For large $T$, this would have negligible effect on the overall performance.} $\bs{s}_\mrm{C}$ is a $TK\times 1$ vector stacking  up the transmitted symbol vectors $\bs{s}[t]$ at $t=1,\dots,T$, $\bs{n}_\mrm{C}$ is the corresponding $(T+1)M\times 1$ AWGN vector, and $\bs{\mathcal{H}}$ is the resulting $(T+1)M\times TK$ ISI channel matrix, which considers simultaneously the space and the time domain. From \eqref{eq:ISI_yt2}, we may express $\bs{\mathcal{H}}$ as
\begin{equation}\label{eq:ISIch}
    \bs{\mathcal{H}}= \begin{bmatrix}
    \bs{H}_0 & \bs{0} & \cdots & \bs{0} \\
    \widetilde{\bs{H}}(\bs{\Theta}) & \bs{H}_0 & \ddots & \vdots \\
    \bs{0} & \widetilde{\bs{H}}(\bs{\Theta}) & \ddots & \bs{0} \\
     \vdots & \ddots & \ddots  & \bs{H}_0 \\
     \bs{0} & \cdots  & \bs{0} & \widetilde{\bs{H}}(\bs{\Theta}) \\
    \end{bmatrix} \hspace{-0.5em}.
\end{equation}
In order to achieve simultaneous orthogonality  in time an space, the ISI channel should fulfill
\begin{equation}\label{eq:ch_orth}
    \bs{\mathcal{H}}^\mrm{H}\bs{\mathcal{H}} = \beta \mathbf{I}_{TK}.
\end{equation}
Note that fulfilling \eqref{eq:ch_orth} allows transforming \eqref{eq:ISI_y_vec} into a set of $TK$ AWGN channels by simply applying MRC to $\bs{y}_\mrm{C}$. We may also relax \eqref{eq:ch_orth} by allowing the right-hand side (RHS) to be an arbitrary diagonal matrix instead of a scaled identity. However, this would lead to having different channel capacity for different UEs, or under different time-slots, which is generally not as desirable as ensuring a fair rate distribution (see \cite{rs_orth_journ}). Nevertheless, the results from this section have trivial extension to that case. Using \eqref{eq:ISIch}, we may operate the left-hand side (LHS) of \eqref{eq:ch_orth} to reach
\begin{equation}
    \bs{\mathcal{H}}^\mrm{H}\bs{\mathcal{H}} = \begin{bmatrix}
    \bs{G}(\bs{\Theta}) &  \bs{Z}^\mrm{H}(\bs{\Theta})  & \bs{0} & \cdots & \bs{0} \\
    \bs{Z}(\bs{\Theta}) & \bs{G}(\bs{\Theta}) & \bs{Z}^\mrm{H}(\bs{\Theta}) & \ddots & \vdots \\
    \bs{0} & \bs{Z}(\bs{\Theta}) & \bs{G}(\bs{\Theta})  & \ddots & \bs{0}\\
    \vdots & \ddots & \ddots & \ddots & \bs{Z}^\mrm{H}(\bs{\Theta})\\
     \bs{0}& \cdots & \bs{0} &\bs{Z}(\bs{\Theta})  & \bs{G}(\bs{\Theta})
    \end{bmatrix},
\end{equation}
where we have defined the $K\times K$ matrices
\begin{subequations}\label{eq:G_Z_mats}
\begin{equation}\label{eq:G_mats}
    \bs{G}(\bs{\Theta}) = \bs{H}_0^\mrm{H} \bs{H}_0+\widetilde{\bs{H}}^\mrm{H}(\bs{\Theta})\widetilde{\bs{H}}(\bs{\Theta})
\end{equation}
\begin{equation}\label{eq:Z_mats}
    \bs{Z}(\bs{\Theta}) = \bs{H}_0^\mrm{H}\widetilde{\bs{H}}(\bs{\Theta}).
\end{equation}  
\end{subequations}
The orthogonality constraint, given in \eqref{eq:ch_orth}, can then be translated into the following conditions
\begin{subequations}\label{eq:orth_G_Z}
    \begin{equation}\label{eq:orth_G}
        \bs{G}(\bs{\Theta}) = \beta \mathbf{I}_K
    \end{equation}
    \begin{equation}\label{eq:orth_Z}
        \bs{Z}(\bs{\Theta}) = \bs{0}_{K \times K}.
    \end{equation}
\end{subequations}
Taking into account the expression for $\bs{Z}(\bs{\Theta})$ in \eqref{eq:Z_mats}, in order to fulfill \eqref{eq:orth_Z} $\bs{\Theta}$ should be selected such that the columns of $\widetilde{\bs{H}}(\bs{\Theta})$ fall in the null-space of $\bs{H}_0^\mrm{H}$. Since $\bs{\Theta}$ is unrestricted, $\widetilde{\bs{H}}(\bs{\Theta})$ may be arbitrarily selected to fulfill said restriction. Let us express the singular value decomposition (SVD) of $\bs{H}_0$ as
\begin{equation}
    \bs{H}_0 = \bs{U}_0 \begin{bmatrix} \bs{\Lambda}_0^{\frac{1}{2}}\\ \bs{0}
    \end{bmatrix}\bs{V}_0,
\end{equation}
where $\bs{U}_0$ and $\bs{V}_0$ are the respective $M\times M$ and $K \times K$ unitary matrices, and $\bs{\Lambda}_0^{\frac{1}{2}}$ is the $K\times K$ diagonal matrix with the singular values along the diagonal. We may then select
\begin{equation}\label{eq:Ht_null}
    \widetilde{\bs{H}}(\bs{B})\triangleq \widetilde{\bs{H}}\big(\bs{\Theta}_\mrm{s}(\bs{B})\big) = \bs{U}_0 \begin{bmatrix}
        \bs{0}_{K \times K}\\ 
        \bs{B}
    \end{bmatrix},
\end{equation}
where $\bs{B}$ corresponds to an $(M-K)\times K$ matrix which may be arbitrarily selected. Note that \eqref{eq:Ht_null} generalizes any $M\times K$ matrix falling in the null-space of $\bs{H}_0^\mrm{H}$. Moreover, $\bs{\Theta}_\mrm{s}(\bs{B})$ may be trivially obtained from \eqref{eq:Ht_null} by inverting $\bs{H}_1$ and $\bs{H}_2$ as
\begin{equation}\label{eq:thet_s}
    \bs{\Theta}_\mrm{s}(\bs{B}) = \bs{H}_1^{-1} \widetilde{\bs{H}}(\bs{B}) \bs{H}_2^\dagger,
\end{equation}
where $\bs{H}_2^\dagger$ corresponds to the left pseudo-inverse of $\bs{H}_2$---which may have multiple solutions. Alternative solutions may consider the inversion of the cascaded channel $(\bs{H}_2^\mrm{T} \otimes \bs{H}_1)$ associated to the vectorized version of \eqref{eq:thet_s}.\footnote{The reader may refer to \cite{rs_orth_journ} for detailed results on the connection between $\bs{\Theta}$ and $\widetilde{\bs{H}}(\bs{\Theta})$ under architectures leading to various resctrictions on $\bs{\Theta}$.} 

Selecting $\widetilde{\bs{H}}(\bs{\Theta})$ by \eqref{eq:Ht_null} allows to fulfill the restriction \eqref{eq:orth_Z}, while maintaining the remaining DoF in the selection of $\bs{B}$. Next, we would like to determine $\bs{B}$ such that \eqref{eq:orth_G} can be simultaneously fulfilled, leading to channel orthogonalization. Considering \eqref{eq:Ht_null} and \eqref{eq:G_mats}, equation \eqref{eq:orth_G} leads to
\begin{equation}\label{eq:orth_B}
    \bs{B}^\mrm{H}\bs{B} = \beta \mathbf{I}_K-\bs{H}_0^\mrm{H} \bs{H}_0.
\end{equation}
The previous equation is only solvable whenever we can match the rank of the two sides. For an generic $\bs{H}_0$---i.e., excluding those in a set of measure 0---and assuming $M\geq K$, the right-hand side is of full-rank $K$ with probability 1. Thus, we also need the LHS to be of full-rank $K$, which leads to the orthogonalization condition
\begin{equation}\label{eq:cond_orth}
    M \geq 2K,
\end{equation}
where we considered that $\mrm{rank}(\bs{B}^\mrm{H}\bs{B})\leq \min(M-K,K)$. Assuming \eqref{eq:cond_orth} is fulfilled, \eqref{eq:orth_B} can be solved by selecting
\begin{equation}\label{eq:B_sol}
    \bs{B}(\widetilde{\bs{U}}) = \widetilde{\bs{U}} \big( \beta \mathbf{I}_K-\bs{\Lambda}_0 \big)^{\frac{1}{2}}\bs{V}_0^\mrm{H},
\end{equation}
where $\widetilde{\bs{U}}$ is a semi-unitary matrix of dimension $(M-K)\times K$, i.e., $\widetilde{\bs{U}}^\mrm{H}\widetilde{\bs{U}}=\mathbf{I}_K$, which may be arbitrarily selected. The result in \eqref{eq:B_sol} corresponds to the selection $\bs{B}=(\beta \mathbf{I}_K-\bs{H}_0^\mrm{H} \bs{H}_0)^{\frac{1}{2}}$, but we have expressed it in terms of its SVD to outline the DoFs left after enforcing the orthogonality constrain, which are captured in the selection of $\widetilde{\bs{U}}$. In fact, these DoFs can be used to reduce the power requirements for performing the RRTx process, as will be discussed in the next section. On the other hand, the selection \eqref{eq:B_sol} also leads to an extra orthogonalization condition given by
\begin{equation}\label{eq:cond_orth_bet}
    \beta \geq \lambda_{0,\max},
\end{equation}
where $\lambda_{0,\max}$ corresponds to the greatest eigenvalue of $\bs{H}_0^\mrm{H}\bs{H}_0$---i.e., the greatest diagonal element of $\bs{\Lambda}_0$. This comes from avoiding square roots of negative diagonal elements in \eqref{eq:B_sol}, which would not allow to fulfill \eqref{eq:orth_B} since the resulting matrix should be positive semi-definite from its definition. The condition \eqref{eq:cond_orth_bet} may be understood as the minimum channel gain required to be able to compensate the direct channel, which may be matched to a minimum power requirement for the RRTx processing.

\section{Power Minimization}\label{section:Pmin}
In Section~\ref{section:orth}, we have shown how to achieve perfect orthogonalization of the ISI channel associated to the RRTx process implemented by the low-power panels. Condition \eqref{eq:cond_orth} delimits the possibility to perform channel orthogonalization in this framework, so let us assume this condition is fulfilled---which is reasonable in typical LIS scenarios where ${M\gg K}$. Let us also assume that $\beta$ is selected to fulfill \eqref{eq:cond_orth_bet}, which may be achieved by adjusting the amplification in the RRTx process. As previously mentioned, the orthogonalization restriction still allows for DoFs which may be suitably exploited to enforce desirable properties on the RRTx processing. We will next study how to make use of these DoFs. 

A major goal of the proposed scheme is to increase the overall energy efficiency by operating some LIS-panels in low-power mode. Hence, it is reasonable to use the DoFs available from the channel orthogonalization to minimize the power of the processing applied at the low-power panels. Said power may be measured in terms of the squared Frobenius norm of the RRTx processing, leading to the following optimization
\begin{equation}\label{eq:norm_min}
\begin{aligned}
    \min_{\widetilde{\bs{U}}}&\; \Vert \bs{\Theta}_\mrm{s}(\widetilde{\bs{U}}) \Vert^2_\mrm{F}\\
    \mrm{s.t.}&\; \widetilde{\bs{U}}^\mrm{H}\widetilde{\bs{U}} =  \mathbf{I}_K,
\end{aligned}
\end{equation}
where we have $\bs{\Theta}_\mrm{s}(\widetilde{\bs{U}})\triangleq \bs{\Theta}_\mrm{s}\big(\bs{B}(\widetilde{\bs{U}})\big)$, i.e., the RRTx processing is obtained from \eqref{eq:thet_s}, using \eqref{eq:Ht_null} and \eqref{eq:B_sol}. Note that \eqref{eq:norm_min} is also relevant for increasing the validity of Assumption~1, since $\Vert \bs{\Theta}_\mrm{s}(\widetilde{\bs{U}}) \Vert^2_\mrm{F}$ directly scales the power of the noise associated to the RRTx process reaching \eqref{eq:ISI_yt}. We may simplify the objective function in \eqref{eq:norm_min} by
\begin{equation}\label{eq:fro_norm}
\Vert\bs{\Theta}_\mrm{s}(\widetilde{\bs{U}}) \Vert^2_\mrm{F} \triangleq \mrm{tr}\big(\bs{\Theta}^\mrm{H}_\mrm{s}(\widetilde{\bs{U}})\bs{\Theta}_\mrm{s}(\widetilde{\bs{U}})\big) \!= \mrm{tr}\big(\widecheck{\bs{H}}_1\widetilde{\bs{U}} \widecheck{\bs{H}}_2\widecheck{\bs{H}}_2^\mrm{H} \widetilde{\bs{U}}^\mrm{H} \widecheck{\bs{H}}_1^\mrm{H}\big),
\end{equation}
where we have defined 
\begin{subequations}
\begin{equation}
    \widecheck{\bs{H}}_1 \triangleq \bs{H}_1^{-1} \bs{U}_0\begin{bmatrix}
        \bs{0}_{K \times (M-K)}\\ 
        \mathbf{I}_{M-K}
        \end{bmatrix}
\end{equation}
\begin{equation}
    \widecheck{\bs{H}}_2 \triangleq \big( \beta \mathbf{I}_K-\bs{\Lambda}_0 \big)^{\frac{1}{2}}\bs{V}_0^\mrm{H}\bs{H}_2^\dagger.
\end{equation}   
\end{subequations}
Since we want to solve \eqref{eq:norm_min}, we may select without loss $\bs{H}_2^\dagger$ as the Moore-Penrose inverse \cite{matrix_an}, since any other selection corresponds to adding a matrix in the left null-space of $\bs{H}_2$, which would give a positive additive term in $\Vert\bs{\Theta}_\mrm{s}(\widetilde{\bs{U}}) \Vert^2_\mrm{F}$, i.e., leading to a suboptimal result.

\renewcommand{\baselinestretch}{.95} 

The minimization problem defined in \eqref{eq:norm_min} is non-convex due to the semi-unitary constraint on $\widetilde{\bs{U}}$. However, this constraint corresponds to restricting the search space to the Stiefel manifold $\mathcal{S}\big(M-K,K \big)$, which, when equipped with a Riemannian metric, gives a Riemannian manifold. We may then employ the Riemannian structure to solve \eqref{eq:norm_min} within the Stiefel manifold defined from its restrictions. Specifically, we will consider the results from \cite{traian}, which studies the Riemannian struture of the unitary group under the bi-invariant Riemannian metric induced by the inner product $\langle \bs{U}, \bs{V}\rangle=\frac{1}{2}\mrm{tr}(\bs{U} \bs{V}^\mrm{H})$. Note that the particularization of results from the unitary group $\mathcal{U}(M-K)$ to the Stiefel manifold $\mathcal{S}\big(M-K,K \big)$ can be done by considering that any $\widetilde{\bs{U}}\in\mathcal{S}\big(M-K,K \big)$ corresponds to a projection of a unitary matrix $\bs{U}\in \mathcal{U}(M-K)$ given by $\widetilde{\bs{U}}=\bs{U}\begin{bmatrix}
    \mathbf{I}_K & \bs{0}
\end{bmatrix}^\mrm{T}$.

Considering the Riemannian structure studied in \cite{traian} for the unitary group, we may express the Riemannian gradient at an arbitrary point $\bs{U}\in \mathcal{U}(M-K)$ as
\begin{equation}\label{eq:Riem_grad}
    \tilde{\nabla}\mathcal{J}(\bs{U}) = \bs{\Gamma}_{\mathcal{J}}(\bs{U})-\bs{U}\bs{\Gamma}^\mrm{H}_{\mathcal{J}}(\bs{U})\bs{U},
\end{equation}
where $\bs{\Gamma}_{\mathcal{J}}(\bs{U})\triangleq\frac{\partial\mathcal{J}(\bs{U})}{\partial \bs{U}^*}$ corresponds to the Euclidean gradient of the objective function $\mathcal{J}(\bs{U})$.\footnote{We consider differentiating only over the frame $\partial \bs{U}^*$, which does not span by itself the whole complex tangent space. However, since our analysis depends upon finding stationary points, while a stationary point under said frame equivalently corresponds to a stationary point under the frame $\partial \bs{U}$ \cite{complex_stationary} (and vice-versa), we have no loss in differentiating only over $\bs{U}^*$.} Since we are interested in solving \eqref{eq:norm_min}, the objective function corresponds to $\mathcal{J}(\bs{U})=\big\Vert \bs{\Theta}_\mrm{s}\big(\bs{U}\begin{bmatrix}
    \mathbf{I}_K & \bs{0}
\end{bmatrix}^\mrm{T}\big) \big\Vert^2_\mrm{F}$. 
We may note that, since the objective function $\mathcal{J}(\bs{U})$ does not depend on the last $(M-2K)$ columns of $\bs{U}$, the respective entries in $\bs{\Gamma}_{\mathcal{J}}(\bs{U})$ are $0$. Through basic optimization arguments \cite{boyd}, the solution to \eqref{eq:norm_min} should be given either by a stationary point or by a boundary point. When particularizing the gradient to the underlying Riemannian structure, all the points now correspond to interior points, so the solution to \eqref{eq:norm_min} should be exclusively given by stationary points, i.e., such that the Riemannian gradient in \eqref{eq:Riem_grad} vanishes. The stationary points may then be found by solving the equation $\tilde{\nabla}\mathcal{J}(\bs{U})=0$. We then note that the $(M-2K)$ columns of $\bs{U}$ not appearing in $\widetilde{\bs{U}}$ have no impact on the stationary points of \eqref{eq:Riem_grad} since an equivalent equation to the previous one is obtained by multiplying by $\bs{U}^\mrm{H}$ both sides from the right, so that in the resulting equation the respective columns of $\bs{U}$ appear multiplied by the $0$ entries in $\bs{\Gamma}_{\mathcal{J}}(\bs{U})$. We thus reach the equivalent equation
 \begin{equation}\label{eq:stat_pt}
     \bs{\Gamma}_{\mathcal{J}}(\widetilde{\bs{U}})\widetilde{\bs{U}}^\mrm{H} = \widetilde{\bs{U}}\bs{\Gamma}^\mrm{H}_{\mathcal{J}}(\widetilde{\bs{U}}),
 \end{equation}
where the Euclidean gradient $\bs{\Gamma}_{\mathcal{J}}(\widetilde{\bs{U}})\triangleq\frac{\partial \mathcal{J}(\widetilde{\bs{U}})}{\partial \widetilde{\bs{U}}^*}$ corresponds to the non-zero part of $\bs{\Gamma}_{\mathcal{J}}(\bs{U})$. From \eqref{eq:fro_norm}, we may get \cite{matrix_diff}
\begin{equation}
    \bs{\Gamma}_{\mathcal{J}}(\widetilde{\bs{U}})= \widecheck{\bs{G}}_1 \widetilde{\bs{U}}\widecheck{\bs{G}}_2,
\end{equation}
 where we have defined $\widecheck{\bs{G}}_1=\widecheck{\bs{H}}_1^\mrm{H}\widecheck{\bs{H}}_1$ and $\widecheck{\bs{G}}_2=\widecheck{\bs{H}}_2\widecheck{\bs{H}}_2^\mrm{H}$. Substituting in \eqref{eq:stat_pt} gives the following matrix equation
 \begin{equation}\label{eq:stat_eq}
     \widecheck{\bs{G}}_1 \widetilde{\bs{U}}\widecheck{\bs{G}}_2\widetilde{\bs{U}}^\mrm{H} = \widetilde{\bs{U}}\widecheck{\bs{G}}_2^\mrm{H} \widetilde{\bs{U}}^\mrm{H}\widecheck{\bs{G}}_1^\mrm{H}.
 \end{equation}
The following lemma characterizes the unique set of solutions to \eqref{eq:stat_eq} for any full-rank $\widecheck{\bs{G}}_1$ and $\widecheck{\bs{G}}_2$ with distinct eigenvalues. Note that this condition is fulfilled with probability 1 for generic $\bs{H}_1$ and $\bs{H}_2$ matrices. However, whenever it is not fulfilled, the solution to \eqref{eq:norm_min} described next may be extended by analogously exploiting the extra DoFs.
\begin{lem}\label{lem:sol_stat_eq}
Consider two arbitrary positive-definite matrices $\widecheck{\bs{G}}_1$ and $\widecheck{\bs{G}}_2$, each of them containing distinct eigenvalues, and where we may assume without loss that the dimension of $\widecheck{\bs{G}}_1$ is larger than that of $\widecheck{\bs{G}}_2$. The solutions to \eqref{eq:stat_eq} with semi-unitary constraint on $\widetilde{\bs{U}}$ fulfill
\begin{equation}\label{eq:sol}
    \widetilde{\bs{U}} = \widecheck{\bs{U}}_1 \bs{P}_1\begin{bmatrix}
        \bs{\Lambda}_{\bs{\varphi}}\\
        \bs{0}
    \end{bmatrix}\bs{P}_2 \widecheck{\bs{U}}_2^\mrm{H},
\end{equation}
where $\widecheck{\bs{U}}_1$ and $\widecheck{\bs{U}}_2$ are the unitary matrices from the eigenvalue decomposition (with decreasing eigenvalues) of $\widecheck{\bs{G}}_1$ and $\widecheck{\bs{G}}_2$, respectively, $\bs{P}$ is an arbitrary permutation matrix, and $\bs{\Lambda_{\bs{\varphi}}}$ is a diagonal matrix with unit modulus entries along the diagonal.
\begin{skproof}
The proof consists of expressing $\widehat{\bs{G}}_1$ and $\widehat{\bs{G}}_2$ through their eigenvalue decompositions, and absorbing in $\widetilde{\bs{U}}$ their respective unitary matrices of eigenvectors by having $\widetilde{\bs{U}}=\widecheck{\bs{U}}_1 \widetilde{\widetilde{\bs{U}}} \widecheck{\bs{U}}_2^\mrm{H}$. The resulting equation is then solved by any $\widetilde{\widetilde{\bs{U}}}$ that simultaneously diagonalizes $\widecheck{\bs{\Lambda}}_1$ and $\widecheck{\bs{\Lambda}}_2$, corresponding to the respective diagonal eigenvalue matrices of $\widecheck{\bs{G}}_1$ and $\widecheck{\bs{G}}_2$. Assuming $\widecheck{\bs{G}}_1$ has greater dimension than $\widecheck{\bs{G}}_2$, and both matrices have distinct eigenvalues, the solutions are then trivially captured by \eqref{eq:sol}. A complete proof may be included in the extended version.
\end{skproof}
\end{lem}
\renewcommand{\baselinestretch}{.94} 

From Lemma~\ref{lem:sol_stat_eq}, we may substitute \eqref{eq:sol} into \eqref{eq:fro_norm}, giving
\begin{equation}\label{eq:norm_sol}
    \Vert\bs{\Theta}_\mrm{s} \Vert^2_\mrm{F}=\sum_{i=1}^K \check{\lambda}_{1,\pi_1(i)}\check{\lambda}_{2,\pi_2(i)},
\end{equation}
where $\check{\lambda}_{1,i}$ and $\check{\lambda}_{2,i}$ correspond to the $i$th eigenvalue of $\widecheck{\bs{G}}_1$ and $\widecheck{\bs{G}}_2$, respectively, while $\pi_1$ and $\pi_2$ correspond to the ${(M-K)}$- and $K$-element  permutations of associated to $\bs{P}_1$ and $\bs{P}_2$, respectively. From \eqref{eq:norm_sol}, we can note that the diagonal values of $\bs{\Lambda}_{\bs{\varphi}}$ have no impact whatsoever on the objective function since the corresponding phases cancel each other after being multiplied by their conjugates. We may thus select $\bs{\Lambda}_{\bs{\varphi}}=\bs{I}_K$ without loss of generality. On the other hand, since only $K$ of the $M-K$ eigenvalues of $\widecheck{\bs{G}}_1$ appear in \eqref{eq:norm_sol}, we can trivially select $\pi_1$---through the respective $\bs{P}_1$---such that eigenvalues of $\widecheck{\bs{G}}_1$ appearing in \eqref{eq:norm_sol} are the $K$ eigenvalues with lower value. Alternatively, we may equivalently substitute $\widetilde{\bs{U}}_1$ in \eqref{eq:sol}, associated to the eigenvalue decompostion of $\widecheck{\bs{G}}_1$ for increasing eigenvalues, with the respective unitary matrix associated to the eigenvalue decompositon for decreasing eigenvalues and select $\bs{P}_1=\mathbf{I}_{M-K}$. In any case, we may trivially fix $\bs{P}_1$ without loss, and the remaining DoFs are captured exclusively by $\pi_2$ (or correspondingly $\bs{P}_2$). Hence, with the previous considerations, the search space for the global minimum of \eqref{eq:norm_min} is reduced to evaluating the $K!$ possible options for $\pi_2$ in \eqref{eq:norm_sol}, and choosing the one attaining lowest value. However, me may further obtain a closed form optimal selection of $\pi_2$ by considering the following lemma.
\begin{lem}\label{lem:eig_min_perm}
    Considering a finite increasing sequence $\{\check{\lambda}_{1,i}\}_{1\leq i \leq K}$ and an arbitrary permutation $\pi_2$ of another finite increasing sequence $\{\check{\lambda}_{2,i}\}_{1\leq i \leq K}$, the solution to
    \begin{equation}\label{eq:min_eig}
        \min_{\pi_2} \sum_{i=1}^K\check{\lambda}_{1,i}\check{\lambda}_{2,\pi_2(i)}
    \end{equation}
    is given by the permutation $\pi_2^{\mrm{opt}}$ attaining a decreasing sequence $\{\check{\lambda}_{2,\pi_2^{\mrm{opt}}(i)}\}_{1\leq i \leq K}$.
\begin{skproof}
    The proof can be extended from the $K=2$ case, where we can express $\check{\lambda}_{1,1}=\check{\lambda}_{1,2}+\epsilon_1$ and $\check{\lambda}_{2,1}=\check{\lambda}_{2,2}+\epsilon_2$ for some $\epsilon_1,\epsilon_2 > 0$. If we select $\pi_2$ such that $\{\check{\lambda}_{2,\pi_2(i)}\}_{1\leq i \leq 2}$ is increasing instead of dectreasing we get an extra additive term $\epsilon_1 \epsilon_2$, which is positive from the definition of $\epsilon_1$ and $\epsilon_2$. Hence, the optimal choice is to have $\pi_2$ such that $\{\check{\lambda}_{2,\pi_2^\mrm{opt}(i)}\}_{1\leq i \leq 2}$ is decreasing. A formal proof may be included in the extended version.
\end{skproof}
\end{lem}

If we put together Lemmas~\ref{lem:sol_stat_eq} and \ref{lem:eig_min_perm}, as well as our observation on how to fix $\bs{P}_1$, we can obtain the closed form global optimum $\widetilde{\bs{U}}^\mrm{opt}$ to the minimization problem from \eqref{eq:norm_min}---up to some extra DoFs with no impact on $\Vert \bs{\Theta}_\mrm{s}(\widetilde{\bs{U}}) \Vert^2_\mrm{F}$. 

\section{Numerical Results}\label{section:num_res}
Due to the channel power requirement for the orthogonalization condition in \eqref{eq:cond_orth_bet}, the power of the direct channel $\bs{H}_0$, which is linearly related to the maximum eigenvalue of $\bs{H}_0^\mrm{H} \bs{H}_0$, is a limiting  factor to be able to efficiently implement the proposed scheme. Specifically, it is the relation between this power and the power of the channels $\bs{H}_1$ and $\bs{H}_2$ what matters, since these scale the retransmitted signals which are processed to compensate the direct channel. Thus, we model $\bs{H}_0$, $\bs{H}_1$, and $\bs{H}_2$ as normalized IID Rayleigh fading channels \cite{mimo}, where their powers are scaled such that $\eta$ gives the relation between the direct channel and the RRTx channel---similar to the $K$-factor in a Ricean fading channel \cite{mimo}.\footnote{We are aware that the IID Rayleigh channel model may not be accurate when considering P-LIS scenarios since it disregards the inherent spatial correlation of panels \cite{pizzo}, as well as the presence of strong line-of-sight paths \cite{husha_data}. However, it still allows us to assess the main limitations of our methods, which are mainly related to the rank and power of the considered channels, while it detaches them from specific geometries/implementations.}

\begin{figure*}
     \centering
     \begin{subfigure}[b]{0.49\textwidth}
         \centering
         \includegraphics[scale=0.45]{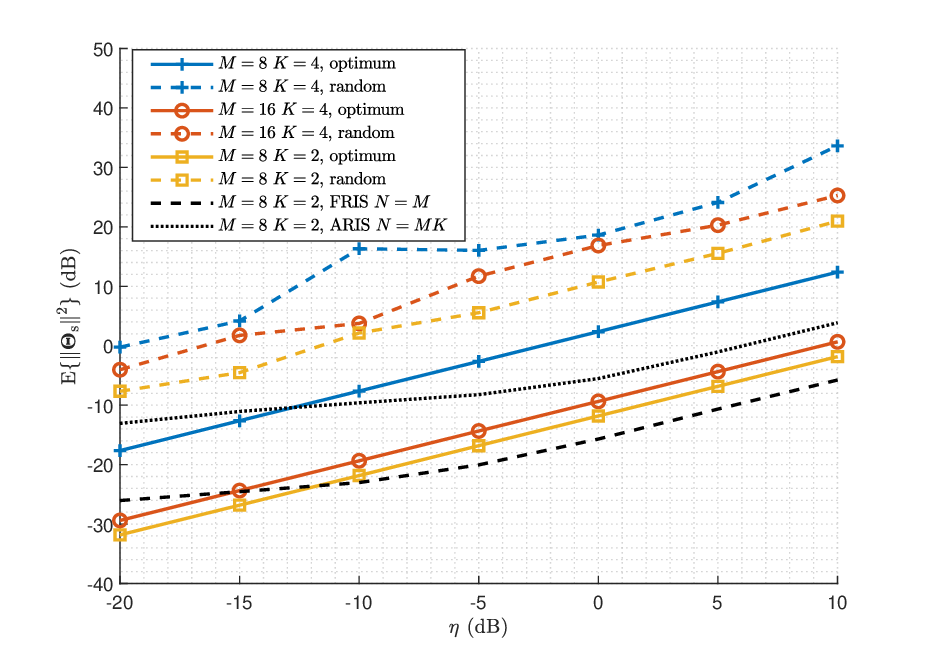}
     \end{subfigure}
     \hfill
     \begin{subfigure}[b]{0.49\textwidth}
         \centering
         \includegraphics[scale=0.45]{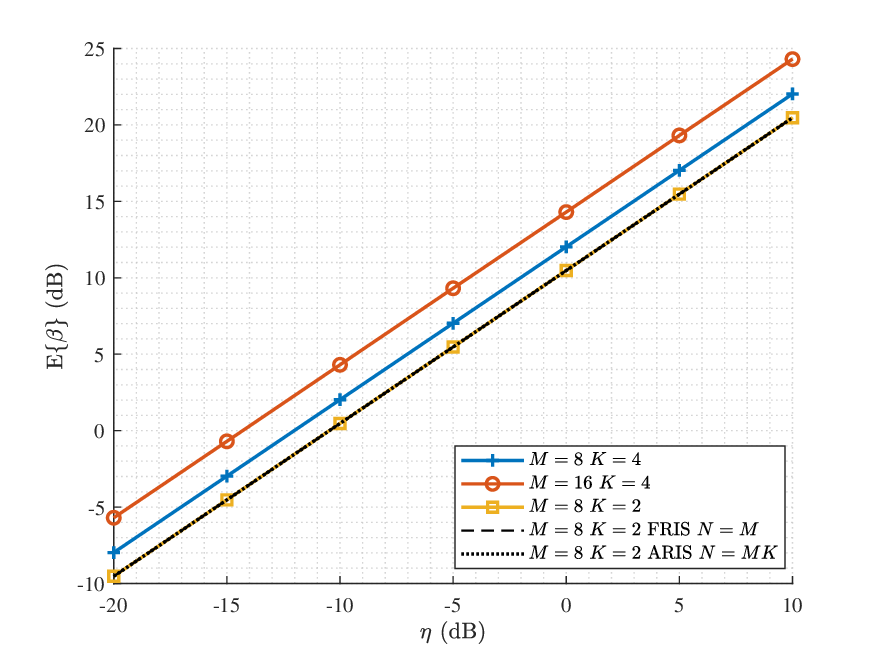}
     \end{subfigure}
     \vspace{-0.25em}
     \caption{RRTx processing power (left) for minimum channel gain per UE (right) with respect to the channel gain ratio.}
        \label{fig:pow_bet}
        \vspace{-1.5em}
\end{figure*}

In Fig.~\ref{fig:pow_bet} (left), we show the average minimum RRTx processing power required to achieve channel orthogonalization for different values of $M$ and $K$. We include results for the optimized DoF selection from Section~\ref{section:Pmin}, as well as for a random selection, i.e., where $\widetilde{\bs{U}}$ in \eqref{eq:B_sol} is isotropically selected, which outlines the importance of our optimization method that allows reducing the required power by roughly $10$ dB. On the other hand, we include as comparison the optimization results from \cite{globecom22}, which solves an analogous optimization using geodesic gradient descent for two reflective surface models, namely amplitude-reconfigurable intelligent surface (ARIS) and FRIS. For equal antenna combinations, the new results outperform those of ARIS, and, for weak direct channels, they also outperform FRIS, while our solution is more efficient since it does not require iterative computation of matrix exponentials. For greater direct channel gains, FRIS still requires lower processing power, but due to its passive nature performance would be largely degraded when considering a multi-carrier system, while the current solution employs robust per-subcarrier BB processing. Moreover, as outlined in \cite{rs_orth_journ}, FRIS may be rather understood as a theoretical concept to upper limit the capabilities of such systems, while our solution still performs close to it. We should also consider the respective channel gain associated to these processing power curves, shown in Fig.~\ref{fig:pow_bet} (right). We may note that the proposed method attains the same minimum  $\beta$ as that of FRIS and ARIS for the same antenna configurations, hinting that a similar rule to \eqref{eq:cond_orth_bet} may apply to such systems, while this was not formally proved yet in \cite{globecom22,rs_orth_journ}.

\begin{figure}
     \centering
     \includegraphics[scale=0.46]{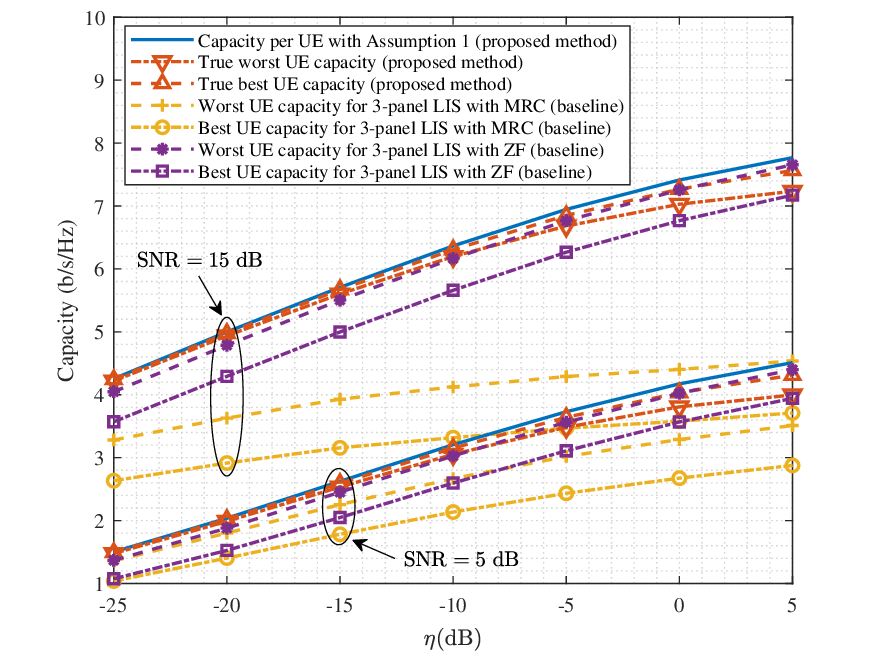}
     \vspace{-0.25em}
        \caption{Ergodic UE capacity with respect to channel gain ratio for $M=8$ antennas per panel and $K=2$ UEs.}
        \label{fig:Cap}
        \vspace{-1.3em}
\end{figure}

Fig.~\ref{fig:Cap} shows the capacity per UE of the proposed method after MRC processing with and without Assumption~\ref{as:noise}, where we include the capacity of the best and the worst UE. We have assumed that the power of the noise added at the low-power panels is the same as that of the noise added at the active panel, i.e., $\tilde{N}_0=N_0$. However, $\tilde{N}_0$ could be actually lower since the low-power panels may operate at low amplification, so that Assumption~\ref{as:noise} may have better accuracy in reality than what is seen from Fig.~\ref{fig:Cap}. In any case, we see that this assumption is still accurate as long as the direct channel gain is not much greater than the gain of RRTx channels, which applies for reasonable blockage of the direct channel, e.g., when the low-power panels connect an outdoor-indoor scenario (or viceversa). We have plotted for comparison the capacity achieved when the 3 LIS-panels actively serves the UEs via joint MRC or zero-forcing (ZF) and under the same channel conditions. We have respectively adjusted the gain of the channels with the power of $\bs{\Theta}$ to ensure fairness. Note that the channel between the LP-Tx panel and the UEs has been ignored in this work since $\bs{H}_1$ connects only the active-LIS panel with the LP-Tx panel, so our approach offers extra robustness to blockages in said channel with respect to the 3-panel fully-active LIS. Our method seems to scale with SNR in the similar to the ZF baseline, which clearly outperforms the MRC baseline. Moreover, even without Asumption~\ref{as:noise}, our method ensures better UE rate fairness than ZF (achieving higher worst UE rate), while it only requires MRC processing in the active panel, posing an important complexity advantage over the ZF approach which requires to coordinate the panels to be able to invert the aggregated channel matrix.\footnote{Assuming $\bs{H}_2$ is compensated (inverted) at the SP-Rx panel, the fronthaul link between the SP-Rx and SP-Tx panels should only support the transmission of a $K\times 1$ vector, while for performing joint ZF the $K\times K$ local channel Gram matrices, as well as the $K\times 1$ MRC-processed vector, should be shared through fronthaul links connecting all 3 panels.}

\section{Conclusions}\label{section:conc}
We have studied channel orthogonalization in a P-LIS scenario with a single active LIS-panel and two low-power LIS panels jointly implementing an RRTx scheme. We have derived the RRTx processing required to perfectly orthogonalize the resulting channel, and characterized the conditions under which this is possible. We have also derived closed-form results on how to utilize the DoFs available from the orthogonality constraint to obtain (globally) minimum RRTx processing power. The numerical results showcase the potential of the methods studied in this work.
\bibliographystyle{IEEEtran}
\bibliography{IEEEabrv,bibliography}

%

\end{document}